# Orbital polarization and third-order anomalous Hall effect in WTe$_2$

Xing-Guo Ye,[*] Peng-Fei Zhu,[*] Wen-Zheng Xu, Zhihao Zang, Yu Ye, and Zhi-Min Liao[†]

*State Key Laboratory for Mesoscopic Physics and Frontiers Science Center for Nano-optoelectronics, School of Physics, Peking University, Beijing 100871, China*

The anomalous Hall effect (AHE) has been extended into the nonlinear regime, where the Hall voltage shows higher-order response to the applied current. Nevertheless, the microscopic mechanism of the nonlinear AHE remains unclear. Here we report the orbital polarization and its induced third-order AHE in few-layer WTe$_2$ flakes. Through angle-dependent electric measurements, it is found that the third-order AHE is quite consistent with the electric field induced polarization of orbital magnetic moment caused by the Berry connection polarizability tensor, which is further directly detected by polar reflective magnetic circular dichroism spectroscopy. The microscopic mechanisms of third-order AHE are analyzed through the scaling law, that is, the opposite orbital magnetic moments (up or down) deflect to opposite directions driven by electric field induced Berry curvature, forming the intrinsic contribution; driven by the Magnus effect of the self-rotating Bloch electrons, the opposite orbital magnetic moments are scattered towards opposite transverse directions, resulting in the skew scattering.

## I. INTRODUCTION

The in-depth investigation of anomalous Hall effect (AHE) in modern condensed matter physics leads to the discovery of Berry curvature, which describes the geometrical properties of Bloch electron wave functions [1–3]. Carriers acquire an anomalous transverse velocity induced by Berry curvature, which is regarded as the intrinsic origin of the AHE [2–4]. Besides the intrinsic contribution to AHE, the extrinsic skew scattering [3] and side jump [4] are related to the spin dependent deflection, that is, the spin up and spin down electrons towards the opposite direction. In magnetic systems, the spin polarization results in a measurable Hall voltage without external magnetic field. In a nonmagnetic system without spin polarization, the spin dependent scattering causes the well-known spin Hall effect [5].

Recently, the AHE has been extended to the higher-order versions. Both second-order and third-order AHEs have been identified theoretically and experimentally [6–9]. The intrinsic contributions of higher-order AHE can be understood within the framework of Berry curvature dipole. The Berry curvature dipole, defined as $D_\alpha = -\frac{1}{\hbar}\int \delta(\varepsilon - \varepsilon_F)\frac{\partial \varepsilon}{\partial k_\alpha}\Omega(k)d^2\mathbf{k}$ (where $\varepsilon$ is the energy, $k_\alpha$ is the wave vector along the $\alpha$ direction, $\varepsilon_F$ is the Fermi energy, and $\Omega$ is Berry curvature), describes the asymmetric distribution of Berry curvature in momentum space [10]. Under the application of an external electric field, polarization of orbital magnetic moment (orbital polarization) can be induced [11,12], following $M \propto \mathbf{E \cdot D}$, where $\mathbf{E}$ and $\mathbf{D}$ are the applied electric field and Berry curvature dipole, respectively. This electric field induced orbital polarization, analog to spin polarization, further lead to the higher-order AHE. The second-order AHE, induced by the inherent Berry curvature dipole, has been found in various systems [7,8,10–18], including bilayer and few-layer WTe$_2$, strained monolayer WSe$_2$, twisted WSe$_2$, corrugated bilayer graphene, TaIrTe$_4$ and MoTe$_2$, etc. Due to the existence of Berry connection polarizability (BCP), nonzero Berry curvature dipole can be generated by an external electric field, resulting in the third-order AHE [9,19], which has been observed in both MoTe$_2$ and WTe$_2$ flakes [9]. Therefore, the direct measurement of the electric field induced orbital polarization is highly desirable and the investigation of the microscopic mechanism of its contribution to the higher-order AHE is necessary.

Here we report the orbital polarization and third-order AHE in time-reversal symmetric WTe$_2$ through both transport and optical measurements. The axial anisotropy of the third-order AHE in WTe$_2$ is found, which is quite consistent with the electric field induced orbital polarization due to the BCP. Through the polar reflective magnetic circular dichroism (RMCD) spectroscopy, it is found that the orbital polarization is parabolic dependent on the electric field. The temperature-dependent transport measurements reveal both intrinsic and skew scattering contributions to the third-order AHE in WTe$_2$ with high carrier mobility in the ultraclean limit. Different from conventional ferromagnetic metals with low mobility, the scaling law for the third-order AHE in WTe$_2$ indicates notable skew scattering, which is attributed to the orbit dependent scattering and the orbital polarization.

## II. EXPERIMENTAL METHOD

The WTe$_2$ flakes with thickness around 5–10 nm were selected to demonstrate the third-order AHE (see Appendix A). In the bulk limit, $T_d$-WTe$_2$ has the space group $Pmn2_1$ possessing a mirror symmetry with the mirror line along the $b$ axis and a glide mirror symmetry with the mirror line along the $a$ axis [20]. The two mirror symmetries recover an inversion in the $ab$ plane, leading to the absence of second-order AHE in WTe$_2$ bulk [6], but the third-order AHE can exist [9]. The WTe$_2$ devices were fabricated with a circular disk electrode structure or a Hall barlike structure. The WTe$_2$ flakes with long, straight edges were selected [7], and the crystalline axis was further confirmed by angle-dependent electric measurements. It is worthwhile to note that the long straight edges of WTe$_2$ do not necessarily correspond to the crystalline axis [21]. The results of four devices, named device S1–S4, are presented in this work. The devices all demonstrate high residual-resistivity ratio $R(273\,\text{K})/R(2\,\text{K})$ ranging $\sim$ 8–19 and electron mobility ranging $2500-5000\,\text{cm}^2/\text{V s}$ (see Appendix B), indicating the high quality of the devices. The temperature dependence of resistivity for these devices (see Appendix B) shows a typical metallic behavior with residual resistivity lower than $10\,\mu\Omega\,\text{cm}$, indicating the ultraclean limit [1].

---

[*]These authors contributed equally to this work.
[†]liaozm@pku.edu.cn

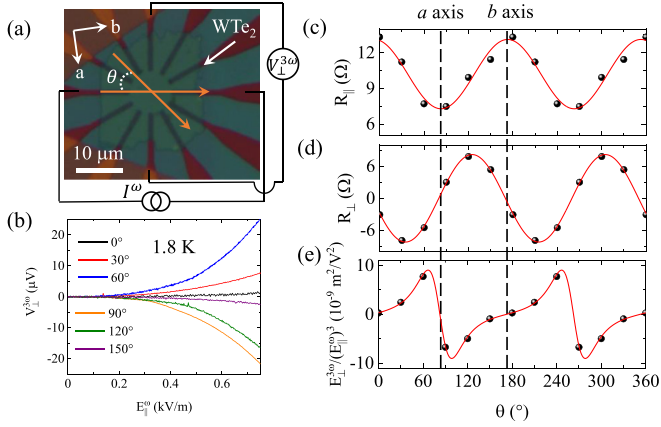

FIG. 1. (a) The optical image of device S1, where the angle $\theta$ is defined. Measurement configuration of the third-order AHE at $\theta = 0°$ is shown. The measurement framework is rotated to collect data at different angles. (b) The angle-dependent third-order AHE in device S1 at 1.8 K. (c)–(e) The angle dependence of the (c) longitudinal ($R_{\parallel}$), (d) transverse ($R_{\perp}$) resistance, and (e) the third-order anomalous Hall signal $\frac{E_{\perp}^{3\omega}}{(E_{\parallel}^{\omega})^3}$, respectively.

## III. RESULTS AND DISCUSSION

Figure 1(a) shows an optical image of device S1, a ten-layer WTe$_2$, with the circular disk electrodes, where the source-drain electrode pairs are perpendicular to the Hall measurement electrode pairs. The third-order AHE was measured at different angle $\theta$, which is defined as the angle between the electric field direction and a baseline of the electrode pair in Fig. 1(a). As shown in Fig. 1(b), the third-harmonic Hall voltage $V_{\perp}^{3\omega}$ measured at 1.8 K demonstrates a clear $\theta$ dependence. The $V_{\perp}^{3\omega}$ shows a cubic dependence on $E_{\parallel}^{3\omega}$, which clearly reveals the third-order nature of the nonlinear Hall response in WTe$_2$. Considering the resistance anisotropy in WTe$_2$, the $E_{\parallel}^{\omega}$ is obtained from $E_{\parallel}^{\omega} = \frac{I^{\omega} R_{\parallel}}{L}$, where $I^{\omega}$ is the applied alternating current, $R_{\parallel}$ is the longitudinal resistance along the current direction, and $L$ is the channel length. Figures 1(c)–1(e) show the angle dependence of the longitudinal ($R_{\parallel}$) and transverse ($R_{\perp}$) resistance and the third-order anomalous Hall signal $\frac{E_{\perp}^{3\omega}}{(E_{\parallel}^{\omega})^3}$, respectively, where $E_{\perp}^{3\omega} = \frac{V_{\perp}^{3\omega}}{W}$ and $W$ is the channel width. The third-order AHE shows the angle dependence following the formula [9]

$$\frac{E_{\perp}^{3\omega}}{(E_{\parallel}^{\omega})^3} \propto \frac{\cos(\theta - \theta_0)\sin(\theta - \theta_0)[(\chi_{22} r^4 - 3\chi_{12} r^2)\sin^2(\theta - \theta_0) + (3\chi_{21} r^2 - \chi_{11})\cos^2(\theta - \theta_0)]}{[\cos^2(\theta - \theta_0) + r\sin^2(\theta - \theta_0)]^3},$$

where $r$ is the resistance anisotropy, $\chi_{ij}$ are elements of the third-order susceptibility tensor, and $\theta_0$ is the angle misalignment between $\theta = 0°$ and the crystalline $b$ axis of WTe$_2$. The fitting curve for this angle dependence is shown by the red line in Fig. 1(e), which yields that the misalignment $\theta_0$ equals 7.4°. The angle dependence of $R_{\parallel}$ and $R_{\perp}$ is also fitted by $R_{\parallel}(\theta) = R_b \cos^2(\theta - \theta_0) + R_a \sin^2(\theta - \theta_0)$ and $R_{\perp}(\theta) = (R_b - R_a)\sin(\theta - \theta_0)\cos(\theta - \theta_0)$, respectively, consistent with the $Pmn2_1$ space group symmetry of WTe$_2$ [8].

The anomalous crystal orientation dependence of the third-order AHE is well explained by the electric field induced orbital polarization due to the BCP. Upon applying an external electric field, the orbital magnetism is induced, following $\mathbf{M} \propto \mathbf{D} \cdot \mathbf{E}$ [22–24]. This current-induced orbital polarization further leads to an AHE in the nonlinear form, including the second-order AHE induced by Berry curvature dipole and the third-order AHE induced by BCP [6–9]. Employing a two-dimensional gapped Dirac model [19], the calculated band structure is shown in Fig. 2(a) (see Appendix C). The BCP $G_{ab}(\mathbf{k})$ is a second-rank tensor, which can generate Berry curvature under the application of an external electric field [19,25], following $\mathbf{\Omega} = \nabla_{\mathbf{k}} \times (\overleftrightarrow{\mathbf{G}} \mathbf{E})$, where the double arrow indicates the second-rank tensor. Figures 2(b) and 2(c) show the BCP-induced Berry curvature under electric fields along the $x$ and $y$ directions, respectively. It is clearly found that this electric field induced Berry curvature shows a dipolelike distribution with nonzero Berry curvature dipole. The magnitude of Berry curvature dipole as a function of the orientation of electric field is shown in Fig. 2(d).

When applying alternating electric field $\mathbf{E}_{\parallel}^{\omega}$ with frequency $\omega$, the induced Berry curvature dipole also holds a frequency $\omega$, satisfying $D^{\omega} \propto E_{\parallel}^{\omega}$. As shown in Fig. 2(e), the $\mathbf{D}^{\omega}$, together with the applied field $\mathbf{E}_{\parallel}^{\omega}$, would further induce orbital polarization $M^{2\omega}$ with frequency $2\omega$, following $M^{2\omega} \propto \mathbf{D}^{\omega} \cdot \mathbf{E}_{\parallel}^{\omega}$ [8,22–24]. This induced orbital polarization would lead to an AHE as a third-order response to $\mathbf{E}_{\parallel}^{\omega}$, the so-called third-order AHE. It is worth noting that as shown in Fig. 1(e), the third-order AHE vanishes when the applied electric field is along the high-symmetry axis of WTe$_2$ ($a$ or $b$ axis). This is consistent with the fact that there is no polarization of orbital magnetic moment as the Berry curvature dipole is perpendicular to the electric field [Figs. 2(b) and 2(c)]. Such strong anisotropic behavior of the observed third-order AHE is quite consistent with the relationship between the induced Berry curvature dipole and the direction of applied electric field in WTe$_2$, as illustrated in Fig. 2(d).

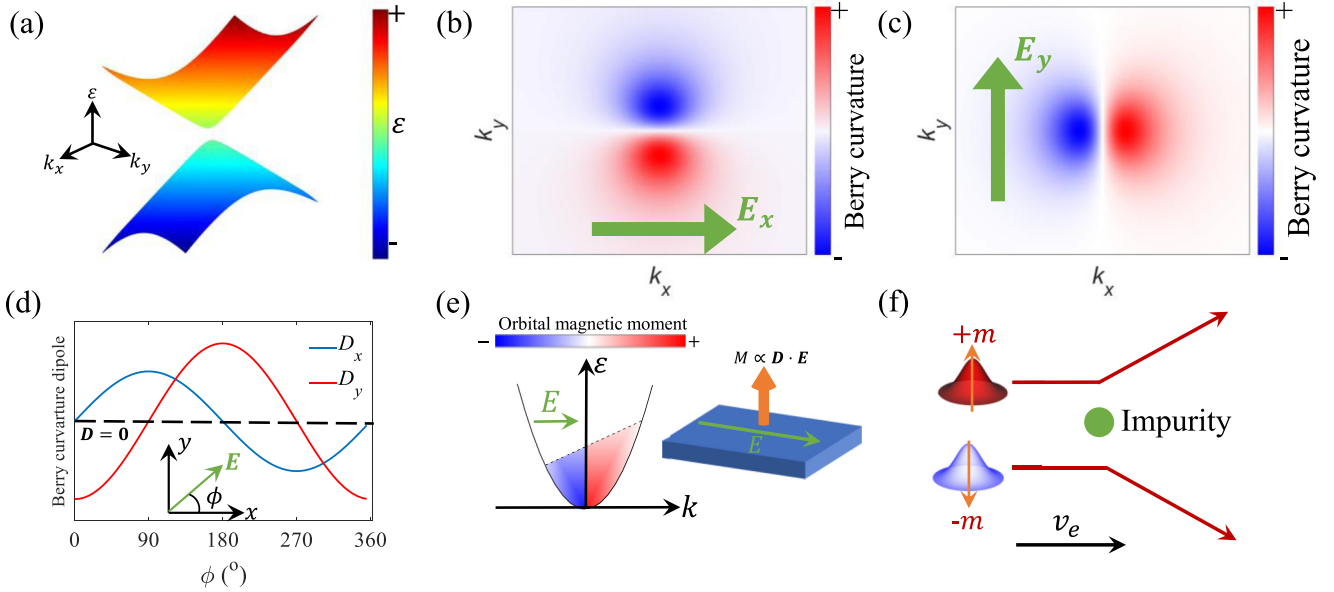

FIG. 2. (a) Band structure of 2D Dirac model with $v_x = 1 \times 10^6$ m/s, $v_y = 0.8 v_x$, $\omega = 0.35 v_x$, and $\Delta = 2$ meV. (b),(c) Electric field induced Berry curvature when electric field is along (b) $x$ axis and (c) $y$ axis. (d) Electric field induced Berry curvature dipole as a function of the orientation of the electric field. (e) Illustration of electric field induced polarization of orbital magnetic moment. (f) Illustration of orbital skew scattering.

To further reveal the electric field induced polarization of orbital magnetic moment, RMCD is measured in WTe$_2$ under the application of an external electric field. RMCD, which measures the differential absorption of left and right circularly polarized light induced by the out-of-plane magnetization of the sample (parallel to the light propagation), is a nondestructive optical method for measuring the magnetism of microsized flakes [26,27]. By exploiting RMCD with the size of a laser spot about 2 $\mu$m in diameter, the dc electric field induced out-of-plane magnetization is directly measured in device S2, a nine-layer WTe$_2$ (see Appendix D). The optical image of device S2 is shown in Fig. 3(a). The third-order AHE is observed in device S2, as shown in Fig. 3(b). Intriguingly, upon applying a dc longitudinal electric field $E_\parallel^{dc}$, a nonzero RMCD signal is observed, as shown in Fig. 3(c). The RMCD signal, which directly reflects the magnitude of out-of-plane magnetization, shows an approximately quadratic dependence on $E_\parallel^{dc}$, consistent with the picture that the third-order AHE results from the electric field induced polarization of orbital magnetic moment following $M \propto \boldsymbol{D} \cdot \boldsymbol{E}$, where the Berry curvature dipole $\boldsymbol{D}$ itself is proportional to the applied field $\boldsymbol{E}$. When the electric field direction is changed by 90° using the electrodes shown in Fig. 3(a), the RMCD measurement results are consistent with the corresponding changes of the third-order AHE (see Appendix D).

We further investigate the polarization of orbital magnetic moment through the scaling law of the third-order AHE. Figures 4(a) and 4(b) show the temperature dependence of the first-harmonic longitudinal voltage and third-order AHE in device S3, an eight-layer WTe$_2$. The scaling relationship between $\frac{E_\perp^{3\omega}}{(E_\parallel^\omega)^3}$ and longitudinal conductivity $\sigma$ is analyzed. The $\sigma$ is changed by varying temperature and calculated by $\sigma = \frac{1}{R_\parallel} \frac{L}{Wd}$, where $d$ is the thickness of the sample. A scaling relationship $\frac{E_\perp^{3\omega}}{(E_\parallel^\omega)^3} \sim \eta + \xi \sigma^2$ is obtained, as shown in Fig. 4(c). Such scaling law is reproducible in other devices (see Appendix E).

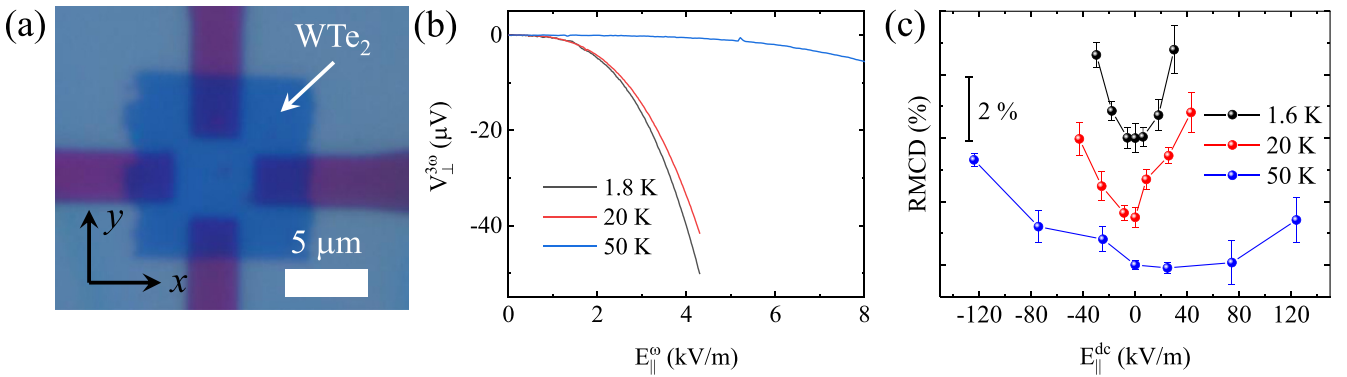

FIG. 3. (a) The optical image of device S2. (b) The third-harmonic Hall voltage in device S2 at various temperatures. (c) The RMCD measurements of device S2 as a function of $E_\parallel^{dc}$. The curves are shifted for clarity.

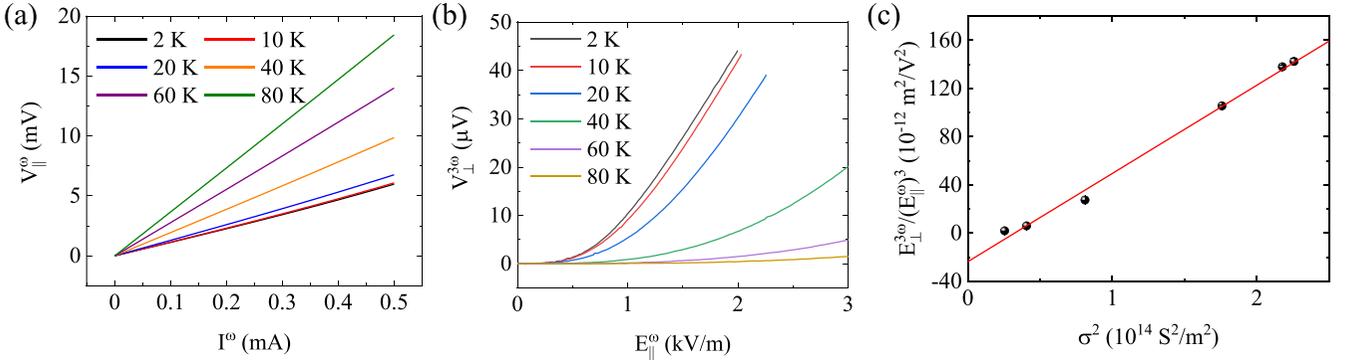

FIG. 4. (a) The first-harmonic longitudinal voltage and (b) the third-order AHE in device S3 at different temperatures. (c) The scaling relationship between $\frac{E_\perp^{3\omega}}{(E_\parallel^\omega)^3}$ and longitudinal conductivity $\sigma$. The conductivity is varied by changing temperature. The temperature range is 2–80 K.

Note that $\sigma \propto \tau$ and $\frac{E_\perp^{3\omega}}{(E_\parallel^\omega)^3} \propto \frac{\chi^{(3)}}{r\sigma}$, where $\tau$ is the scattering time, $r$ is the resistance anisotropy, and $\chi^{(3)}$ is the third-order Hall conductivity. Thus, the linear relationship between $\frac{E_\perp^{3\omega}}{(E_\parallel^\omega)^3}$ and $\sigma^2$ indicates $\chi^{(3)}$ has two contributions, which scales as $\tau^1$ and $\tau^3$, respectively. Note that the electric field induced orbital polarization scales linearly with $\tau$. Therefore, the $\tau^1$ and $\tau^3$ contributions to $\chi^{(3)}$ is quite consistent with the Berry curvature-related anomalous velocity and orbital skew scattering, respectively. This is somehow different with the conventional magnetic metals, where the anomalous Hall conductivity scales as $\tau^0$ for the intrinsic Berry curvature and side jump contributions [1], and as $\tau^2$ for the skew scattering contributions [28,29]. In the third-order AHE, because the magnetization, that is, polarization of orbital magnetic moment, scales as $\tau^1$ itself, the third-order anomalous Hall conductivity $\chi^{(3)}$ scales as $\tau^1$ and $\tau^3$. It is worth noting that the observed $\tau^1$ contribution to $\chi^{(3)}$ may also be induced by the side jump in addition to the intrinsic mechanism. Nevertheless, the high mobility and low residual resistivity in WTe$_2$ suggest the relative weak side jump, which generally dominates in the dirty metals [1]. In addition, the angle dependence of the third-order AHE is well fitted by the model based on the inherent anisotropy analysis in WTe$_2$, which is also unlikely induced by extrinsic impurities. Moreover, the extrinsic effects, such as the thermal effect, thermoelectric effect, mixing with longitudinal voltage, etc., are carefully discussed, which are not likely the origin of the observed third-order AHE (see Appendix F).

It is worth noting that for ordinary linear AHE, the skew scattering is hardly observed in the good metal regime [30]. Here the observed skew scattering in high conductivity non-magnetic WTe$_2$ can be well understood by considering the electric field induced polarization of orbital magnetic moment. Recent theoretical predication by Isobe et al. [31] shows that the Bloch electrons with opposite self-rotating directions are scattered towards opposite transverse directions behaving like the Magnus effect [32,33]. These self-rotating Bloch electrons carry orbital magnetic moments, and the Bloch electrons with opposite orbital magnetic moment suffer asymmetric (skew) scattering, as shown in Fig. 2(f). Since the orbital magnetic moment is polarized under electric field, a nonzero transverse current is able to be induced by the orbital skew

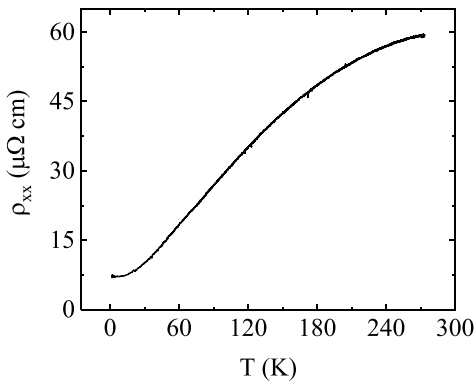

FIG. 5. Temperature dependence of the resistivity of device S3.

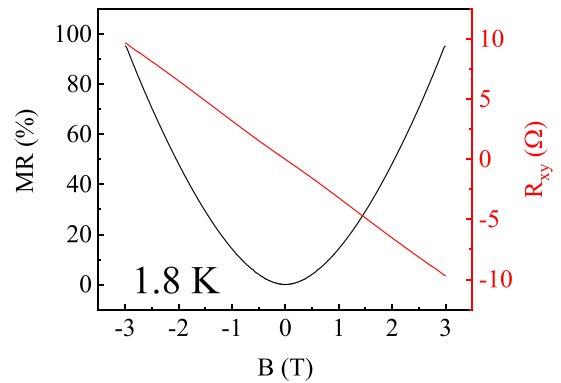

FIG. 6. Magnetotransport of device S1 at 1.8 K.

TABLE I. Summary of thickness, electrode configuration, and RRR of different devices. The RRR is determined as the ratio between the resistivity at 273 and 2 K.

| Device | Thickness | RRR | $\mu_n$ (cm$^2$/V s) | $\mu_p$ (cm$^2$/V s) |
|---|---|---|---|---|
| S1 | 7 nm, 10$L$ | 7.9 | 4985.3 | 3010.1 |
| S2 | 6.3 nm, 9$L$ | 18.3 | NA | NA |
| S3 | 5.6 nm, 8$L$ | 10.8 | NA | NA |
| S4 | 7 nm, 10$L$ | 19.2 | 2520.4 | 2513.9 |

scattering [31], which is also in the third-harmonic response to $\mathbf{E}_\parallel^\omega$.

## IV. CONCLUSIONS

In summary, we have demonstrated the electric field induced orbital polarization and third-order AHE in few-layer WTe$_2$. Notable third-order AHE with strong anisotropy is observed, consistent with the scenario of electric field induced polarization of orbital magnetic moment. Under the application of an external electric field, the BCP tensor would generate Berry curvature with dipolelike distribution, being responsible for the electric field induced orbital polarization. By exploiting RMCD spectroscopy, the electric field induced orbital polarization is directly measured. The scaling law of the third-order AHE further shows that the orbital polarization contributes to the third-order AHE through both intrinsic mechanism and orbital skew scattering. Our work paves the way towards the realization of orbital polarization through electrical means, promising for orbitronics.

## ACKNOWLEDGMENTS

This work was supported by National Key Research and Development Program of China (Grant No. 2018YFA0703703), and National Natural Science Foundation of China (Grants No. 91964201 and No. 61825401).

## APPENDIX A: DEVICE FABRICATION AND MEASUREMENTS

The few-layer WTe$_2$ flakes were obtained from the bulk crystal using a mechanically exfoliating method. Ti/Au electrodes (∼10 nm thick) were prefabricated onto individual SiO$_2$/Si substrates. Exfoliated BN (∼20 nm thick) and few layer WTe$_2$ (around 5–20 nm thick) were sequentially picked up and then transferred onto the Ti/Au electrodes using a polymer-based dry transfer technique [34]. The whole exfoliating and transfer process was done in an argon-filled glove box with O$_2$ and H$_2$O content below 0.01 parts per million to avoid sample degeneration.

The transport measurements were carried out in an Oxford cryostat with a variable temperature insert and a superconducting magnet. The third-harmonic signals were collected by standard lock-in techniques (Stanford Research Systems Model SR830) using an ac excitation with frequency of $\omega =$ 17.777 Hz; the $3\omega$ Hall signals were measured.

## APPENDIX B: TRANSPORT PROPERTIES

In all the devices, the resistivity decreases upon decreasing temperature with a residual resistivity at low temperatures, showing typical metallic behaviors. A typical resistivity-temperature relationship is shown in Fig. 5 measured from device S3. Figure 6 shows the magnetoresistance (MR) and Hall resistance as a function of magnetic field in device S1 at 1.8 K. MR is defined as $\frac{R_{xx}(B)-R_{xx}(0)}{R_{xx}(0)} \times 100\%$. Large, nonsaturated MR with characteristic two-carrier transport is observed, indicating a nearly compensated electron and hole density in WTe$_2$ [35]. Through a semiclassical two-carrier model [36], that is,

$$\rho_{xx} = \frac{1}{e} \frac{n\mu_n + p\mu_p + (n\mu_p + p\mu_n)\mu_n\mu_p B^2}{(n\mu_n + p\mu_p)^2 + (n-p)^2 \mu_n^2 \mu_p^2 B^2},$$

and

$$\rho_{xy} = \frac{1}{e} \frac{(p\mu_p^2 - n\mu_n^2)B + (p-n)\mu_n^2 \mu_p^2 B^3}{(n\mu_n + p\mu_p)^2 + (n-p)^2 \mu_n^2 \mu_p^2 B^2},$$

where $n$ is the electron density, $p$ is the hole density, $\mu_n$ is the electron mobility, and $\mu_p$ is the hole mobility, the carrier density and mobility are estimated. The mobility of all these devices is in the order of $10^3$ cm$^2$/V s. Table I summarizes the thickness, the residual-resistivity ratio (RRR), and the mobility for these devices. Here RRR is determined as the ratio between the resistivity at 273 and 2 K.

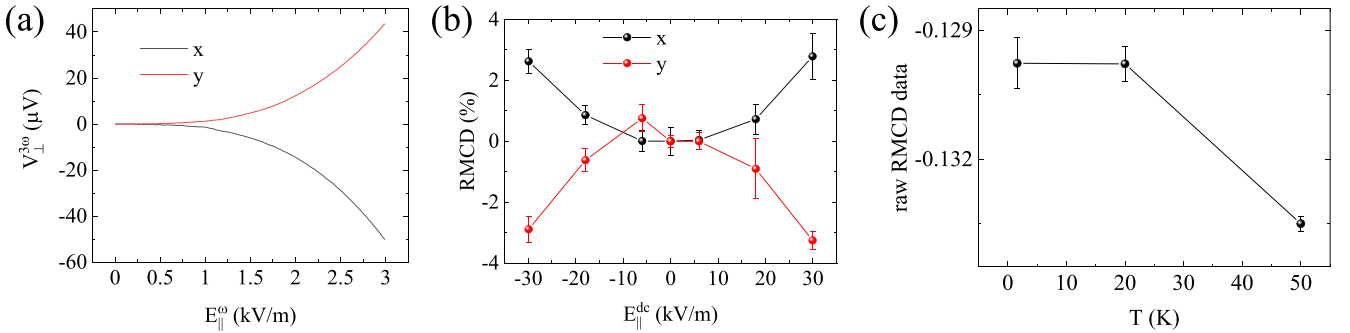

FIG. 7. Electric field direction dependence of the (a) third-order AHE at 1.8 K and (b) RMCD signals at 1.6 K of device S2. (c) Raw data of RMCD signal in device S2 at zero electric field and temperatures from 1.6 to 50 K.

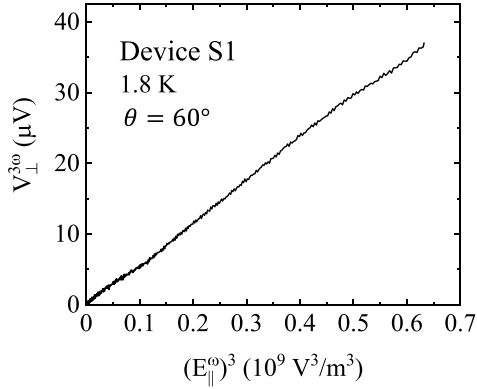

FIG. 8. The third-harmonic Hall voltage $V_\perp^{3\omega}$ as a function of $(E_\parallel^\omega)^3$.

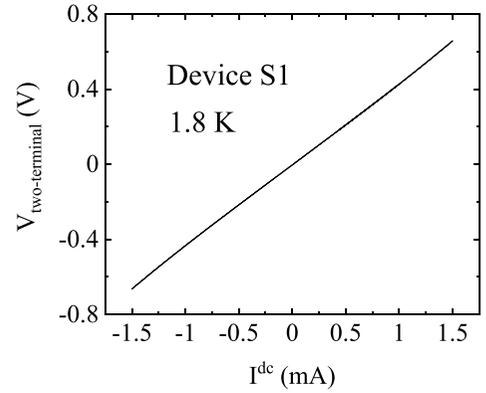

FIG. 10. Two-probe dc measurements of WTe$_2$ in device S1 at 1.8 K, showing linear $I$-$V$ characteristics.

### APPENDIX C: 2D GAPPED DIRAC MODEL

The Hamiltonian of the two-dimensional (2D) gapped Dirac model is $H(\mathbf{k}) = \omega k_x \sigma_0 + v_x k_x \sigma_x + v_y k_y \sigma_y + \Delta \sigma_z$ [19], where $\sigma_0$ is the $2 \times 2$ identity matrix, $\sigma_{x,y,z}$ is the Pauli matrices, $\omega$ describes the tilting of Dirac cone, $v_x$ and $v_y$ describe the anisotropy of Fermi velocity, and $2\Delta$ is the energy gap. This model has two energy bands as $\varepsilon_\pm(\mathbf{k}) = \omega k_x \pm \sqrt{v_x^2 k_x^2 + v_y^2 k_y^2 + \Delta^2}$. The Berry connection polarizability is evaluated, and the field-induced Berry curvature is calculated as $\mathbf{\Omega}^1(\mathbf{k}) = \frac{v_x^2 v_y^2}{2(\sqrt{v_x^2 k_x^2 + v_y^2 k_y^2 + \Delta^2})^5} \mathbf{k} \times \mathbf{E}$. We further calculate the field-induced Berry curvature by $D_\alpha = -\frac{1}{\hbar} \int \delta(\varepsilon - \varepsilon_F) \frac{\partial \varepsilon}{\partial k_\alpha} \Omega(k) d^2k$ ($\varepsilon$ is the energy, $k_\alpha$ is the wave vector along the $\alpha$ direction, $\varepsilon_F$ is the Fermi energy, and $\alpha = x, y$) [7].

### APPENDIX D: POLAR RMCD SPECTROSCOPY

For the RMCD measurements, the He-Ne 633-nm laser was used as the excitation. A chopper and a photoelastic modulator were used to modulate the intensity and polarization of the excitation beam, respectively. The RMCD measurements were carried out in the Attocube closed-cycle cryostat with lowest temperature $\sim 1.6$ K. The reflected beam was collected by the photomultiplier tube and was eventually analyzed by a two-channel lock-in amplifier through the signals of the reflected intensity (at $f_C$) and the RMCD intensities (at $f_{PEM}$).

The dependence of RMCD measurement results on the electric field direction was carried out in device S2. As shown in Fig. 7(a), the third-order AHE shows a similar magnitude but opposite sign along the $x$ and $y$ direction [denoted in Fig. 3(a)]. Similarly, the RMCD measurement holds a consistent result, as shown in Fig. 7(b). This observation suggests that the RMCD signal has the same origin as the third-order AHE.

Besides the orbital polarization, the Joule heating effect may also lead to quadratic dependence of the RMCD signal on the electric field, which, however, can be safely ruled out in our measurements. Figure 7(c) shows the raw RMCD data of device S2 at zero electric field and temperatures from 1.6 to 50 K. It is found that the RMCD background signal weakly decreases upon increasing temperature. Therefore, if the quadratic dependence is induced by Joule heating effect, the RMCD signal should weakly decrease upon applying electric field, which is inconsistent with the results in Fig. 3.

Since it is still debated whether the current induced spin accumulation from the spin Hall effect can be detected optically in metals [37–40], we would like to discuss the difference between the current induced orbital magnetism due to the BCP and the current induced spin accumulation due to the spin Hall effect. In contrast to the in-plane spin direction in a heavy metal surface, the electric field induced orbital polarization in $WTe_2$ is along the out-of-plane direction. Moreover, RMCD is an effective tool to detect out-of-plane magnetization in layered materials [26,27]. On the other hand, the spin Hall effect induced spin accumulation has opposite orientation on the two surfaces of the heavy metals, resulting in the whole system having zero net spin magnetic moments. However, there is a notable orbital magnetization in the $WTe_2$ induced by electric field.

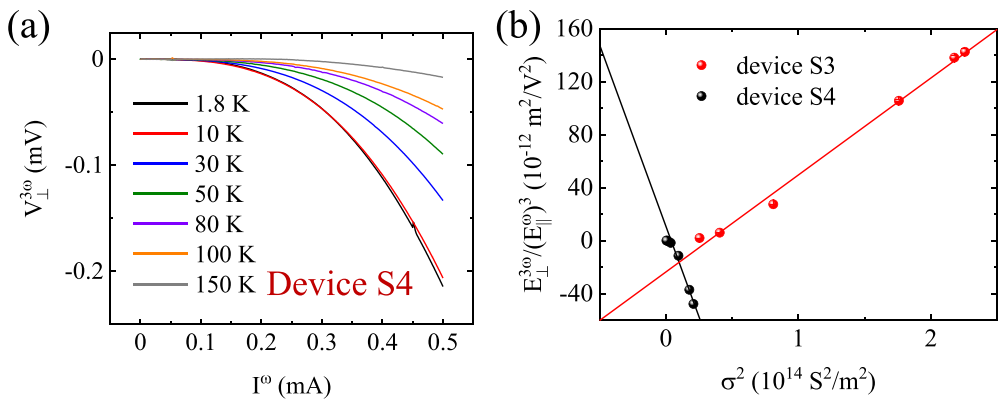

FIG. 9. Scaling law of third-order AHE in device S4. (a) Third-order anomalous Hall voltage $V_\perp^{3\omega}$ at various temperatures. (b) The $\frac{E_\perp^{3\omega}}{(E_\parallel^\omega)^3}$ as a function of $\sigma^2$ in devices S3 and S4. The temperature range for the scaling law is 1.8–80 K.

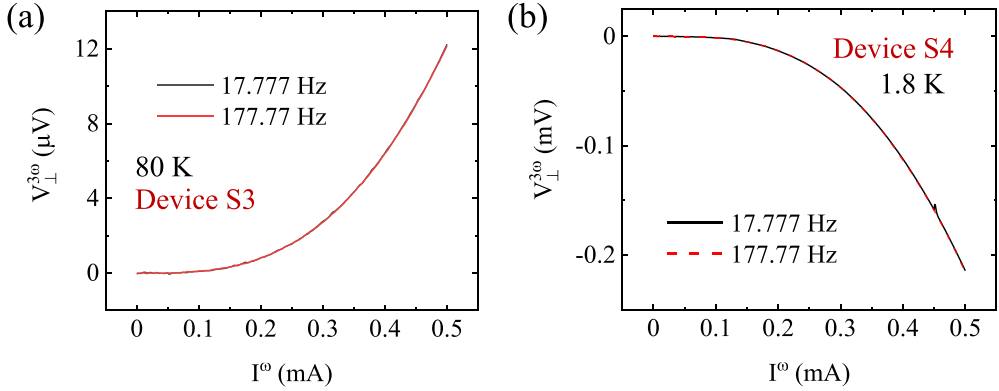

FIG. 11. Frequency dependence of the third-order AHE. The third-order Hall signals at different frequencies (a) in device S3 at 80 K and (b) in device S4 at 1.8 K.

## APPENDIX E: THIRD-ORDER AHE IN DIFFERENT DEVICES

The third-order nature of the AHE is revealed by both the nonzero third-harmonic Hall voltage and the cubic dependence of the third-harmonic Hall voltage on the applied alternating field. As shown in Fig. 8, the third-harmonic Hall voltage $V_\perp^{3\omega}$ shows a linear dependence on $(E_\parallel^\omega)^3$. Moreover, a similar scaling relationship is observed in device S4, as shown in Fig. 9.

## APPENDIX F: OTHER POSSIBLE HIGHER-ORDER RESPONSES

In addition to the third-order AHE induced by intrinsic effect, i.e., the Berry-connection polarizability tensor, there also may exist some extrinsic effects that can induce higher-order transport. Here by control experiments and symmetry analysis, the extrinsic effects are excluded carefully one by one:

*(1) Diode effect.* The Schottky barrier is generally inevitable between the metal electrodes and two-dimensional materials, which would induce an accidental diode with a rectification effect, resulting in high-order transport effect. However, the two-terminal $I$-$V$ characteristics of $WTe_2$ were measured, which show a linear behavior, as shown in Fig. 10, excluding the extrinsic diode effect as the origin of third-order AHE.

*(2) Thermal effect and thermoelectric effect.* The Joule heating can cause the change of sample temperature. As the sample resistance has a notable temperature dependence, it results in a third-order response to the current [41]. On the other hand, the temperature gradient across the sample can also drive a thermoelectric voltage. These effects could be excluded through the angle-dependence measurements of third-order AHE in Fig. 1(e), where the angle dependence of third-order AHE is well fitted by the model based on the internal symmetry analysis of $WTe_2$.

*(3) Transverse resistance $R_\perp$.* The influence of $R_\perp$ on the observed third-order AHE can be ruled out by the angle dependence of the third-order AHE. As shown in Figs. 1(d) and 1(e), $R_\perp$ and the third-order AHE show different angle dependence. For example, at $\theta = 0°$, both $R_\perp$ and the third-order Hall signal $\frac{V_\perp^{3\omega}}{(V_\parallel^\omega)^3}$ show a minimum. However, at $\theta = 30°$, $R_\perp$ shows a maximum while $\frac{V_\perp^{3\omega}}{(V_\parallel^\omega)^3}$ shows a value very close to the minimum (zero). The very different angle dependence between $R_\perp$ and $\frac{V_\perp^{3\omega}}{(V_\parallel^\omega)^3}$ indicates the observed third-order AHE is not produced by $R_\perp$.

*(4) Capacitance coupling effect.* Higher-order transport can be also induced by capacitance coupled within the circuit. To exclude the spurious capacitive coupling effect, frequency-dependent measurements were carried out, as shown in Fig. 11. By applying alternating current with different frequencies (17.777–177.77 Hz), no frequency dependence of the third-order AHE is observed, inconsistent with the spurious capacitive coupling effect, which generally depends strongly on the frequency.

*(5) Mixing from $V_\parallel^{3\omega}$.* The observed third-order Hall signals may be mixed from nonzero third-order longitudinal voltage $V_\parallel^{3\omega}$. To exclude such a possibility, we simultaneously measured the third-harmonic longitudinal and Hall voltage, i.e., $V_\parallel^{3\omega}$ and $V_\perp^{3\omega}$, in device S1 at 1.8 K, as shown in Fig. 12. As shown in Fig. 12(a), in contrast to $V_\perp^{3\omega}$, $V_\parallel^{3\omega}$ shows a relatively small signal. Thus, $V_\perp^{3\omega}$ dominates over $V_\parallel^{3\omega}$. The nonzero $V_\parallel^{3\omega}$ observed here may be induced by thermal effect, as discussed above, or be also induced by the orbital skew scattering, which can induce both transverse and longitudinal nonlinear transport [19]. The origin of the nonzero $V_\parallel^{3\omega}$ needs further investigation. Further, the angle dependence of the $\frac{E_\parallel^{3\omega}}{(E_\parallel^\omega)^3}$ and $\frac{E_\perp^{3\omega}}{(E_\parallel^\omega)^3}$ is also investigated in device S1, as shown in Figs. 12(b) and 12(c). Within a large range of angle $\theta$, $V_\perp^{3\omega}$ dominates over $V_\parallel^{3\omega}$, by direct comparison between Figs. 12(b) and 12(c). Note the angle dependence of third-order AHE $\frac{E_\perp^{3\omega}}{(E_\parallel^\omega)^3}$ is well fitted by the model based on the internal symmetry analysis of WTe$_2$. However, the $\frac{E_\parallel^{3\omega}}{(E_\parallel^\omega)^3}$ and $\frac{E_\perp^{3\omega}}{(E_\parallel^\omega)^3}$ show totally different angle dependence. For example, the $\frac{E_\perp^{3\omega}}{(E_\parallel^\omega)^3}$ is nearly zero when along both $a$ and $b$ axes. Nevertheless, $\frac{E_\parallel^{3\omega}}{(E_\parallel^\omega)^3}$ shows a value close to the maximum when along the $a$ axis and a value close to the minimum along the $b$ axis, as denoted by the black dashed lines in Figs. 12(b) and 12(c). Therefore, through the angle-dependent measurements, the $V_\parallel^{3\omega}$ can be safely excluded as the origin of the observed third-order AHE.

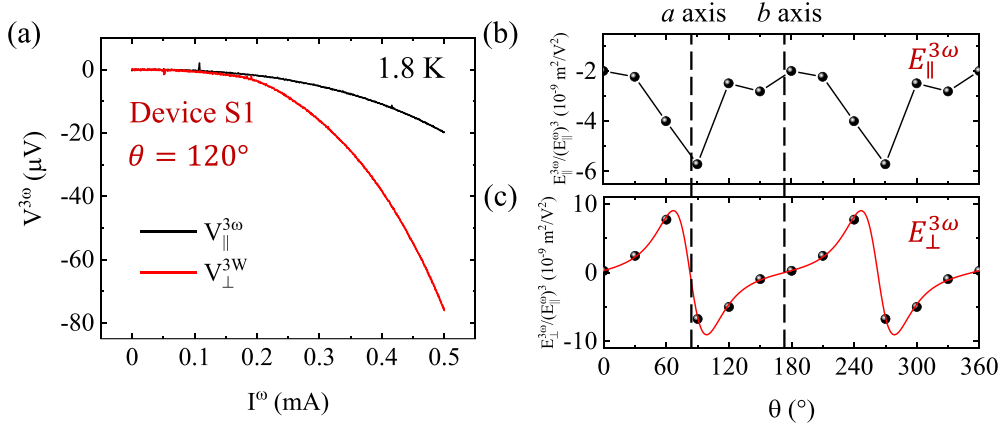

FIG. 12. The third-harmonic longitudinal and Hall voltage, i.e., $V_\parallel^{3\omega}$ and $V_\perp^{3\omega}$, in device S1 at 1.8 K. (a) The simultaneously measured $V_\parallel^{3\omega}$ and $V_\perp^{3\omega}$ at $\theta = 120°$. (b), (c) The angle-dependence of (b) $\frac{E_\parallel^{3\omega}}{(E_\parallel^\omega)^3}$ and (c) $\frac{E_\perp^{3\omega}}{(E_\parallel^\omega)^3}$.